\documentclass{article}

\usepackage{PRIMEarxiv}

\usepackage[utf8]{inputenc} 
\usepackage[T1]{fontenc}    
\usepackage{hyperref}       
\usepackage{url}            
\usepackage{booktabs}       
\usepackage{amsfonts}       
\usepackage{nicefrac}       
\usepackage{microtype}      
\usepackage{lipsum}
\usepackage{fancyhdr}       
\usepackage{graphicx}       
\graphicspath{{media/}}     

\usepackage{comment}
\usepackage{multirow}
\usepackage{tabularx}
\usepackage{amsmath,amssymb}
\usepackage{tikz}
\usetikzlibrary{positioning, fit, calc, shapes, arrows}
\usepackage[underline=false]{pgf-umlsd}
\newcommand{\bloodymess}[7][0]{
  \stepcounter{seqlevel}
  \path
  (#2)+(0,-\theseqlevel*\unitfactor-0.7*\unitfactor) node (mess from) {};
  \addtocounter{seqlevel}{#1}
  \path
  (#4)+(0,-\theseqlevel*\unitfactor-0.7*\unitfactor) node (mess to) {};
  \draw[->,>=angle 60] (mess from) -- (mess to) node [midway, above]
  {#3};

  \if R#5
    \node (#3 from) at (mess from) {\llap{#6~}};
    \node (#3 to) at (mess to) {\rlap{~#7}};
  \else\if L#5
         \node (#3 from) at (mess from) {\rlap{~#6}};
         \node (#3 to) at (mess to) {\llap{#7~}};
       \else
         \node (#3 from) at (mess from) {#6};
         \node (#3 to) at (mess to) {#7};
       \fi
  \fi
}
\usepackage{caption}
\usepackage{epstopdf}
\usepackage{amsmath} 
\usepackage{array}   
\usepackage{setspace} 

\title{Introducing a Novel Secret Key Establishment Method
}

\author{ 
{\hspace{1mm}Luis Adrián Lizama-Pérez}\thanks{https://electronica.usm.cl/personas/lizama-luis/} \\
  Departamento de Electrónica \\
  Universidad Técnica Federico Santa María \\
  Av. Vicuña Mackenna 3939, San Joaquín, Santiago de Chile, 8940897\\
  \texttt{luis.lizamap@usm.cl}\\
 }

\begin{document}
\maketitle

\begin{abstract}
We present a novel approach to secret key establishment that appears to be resistant to currently known quantum cryptanalytic algorithms. This quantum resistance arises because the security of our method does not rely on the difficulty of integer factorization or the discrete logarithm problem. 
Based on the analyses of Alice's public key, the communication exchange between Alice and Bob, and the scenario where Bob behaves as Eve, we can conclude that, even if Eve has access to a quantum computer capable of solving discrete logarithms, she is unable to determine Alice's private key. Additionally, our approach achieves Perfect Forward Secrecy (PFS), ensuring that the security of previously used keys is not compromised by any key that becomes compromised. Notably, our system offers competitive public and private key sizes compared to those currently available.
\end{abstract}


\section{Introduction}

Derived from the advancements in quantum computing and its potential to undermine current cryptographic methods~\cite{shor1994algorithms,barreno2002future,bernstein2009introduction}, the European Telecommunications Standards Institute (ETSI) established the Quantum Secure Cryptography Industry Specification Group (ISG QSC) in 2015. This group aims to assist the industry in addressing the threats posed by quantum computers to existing cryptography. Meanwhile, in 2016, the National Institute of Standards and Technology (NIST) initiated an open competition to foster participation and consensus from both academia and industry for selecting post-quantum cryptography standards in two categories: digital signatures and public key encryption (PKE). While ETSI concentrates on developing standards in Europe with a research-oriented approach to new post-quantum cryptographic methods, NIST focuses on the global standardization process, emphasizing the selection and evaluation of algorithms for inclusion in future standards.

Currently, post-quantum cryptography is still in the research and development phase. The primary approaches under investigation include those based on lattices, Goppa and McEliece codes, hash-based Merkle trees, multivariate cryptography, and elliptic isogeny digital signature schemes. Despite several proposed solutions, there is no consensus on the optimal approach for securing information against quantum attacks. The following are some of the most important post-quantum cryptography techniques:

\begin{itemize}
    \item [1. ] Lattice-based cryptography: This technique is based on mathematical difficulties relating to lattices, which are multidimensional geometric formations. Lattice-based algorithms, such as the GGH and NTRU algorithms, use complex mathematical procedures to encrypt and decrypt data securely. Lattice-based algorithms can be computationally costly, which means they take an extensive amount of resources to operate. This can lead to slower processing times and increased power usage. It is quite a task to improve the computational efficiency of lattice-based algorithms. For example, as compared to other cryptographic algorithms, the computational efficiency of NTRU encryption can be relatively low. This is because the encryption and decryption processes involve costly mathematical operations such as polynomial multiplications and reductions. Furthermore, the size of the keys used in NTRUEncrypt (2048 bits or above) may be relatively large, affecting system efficiency and performance, where NTRU stands for N-degree Truncated Polynomial Ring Units. Other lattice-based approaches are LWE (Learning With Errors), Ring-LWE, and Module-LWE. Cyclotomic rings are used for Ring-LWE, which are special algebraic rings built from roots of unity. Modular lattices, on the other hand, have more sophisticated algebraic structures than ideal lattices. The module learning with errors (MLWE) problem was introduced to solve the shortcomings in both LWE and RLWE. After the third round of evaluation, NIST accepted FALCON (signature), which belongs to the LWE (Learning With Errors) category. CRYSTALS-KYBER (PKE-KEM) in the context of Ring-LWE and CRYSTALS-Dilithium (signature), which is based on Module-LWE.

    \item [2. ] Error-Correction Codes: Goppa Codes and McEliece Codes are error-correcting codes used in post-quantum cryptography. The computational difficulties of addressing a mathematical problem known as the code isomorphism problem are the basis for these codes. When compared to lattice-based algorithms, code-based algorithms, such as McEliece, are often more computationally efficient. When compared to other post-quantum cryptography approaches, code-based algorithms can provide security with smaller keys. However, in comparison to other cryptographic approaches, code-based algorithms often require a larger encrypted message size.

    \item [3. ] Hash-based Merkle Trees: Hash functions are cryptographic techniques that transform input data into a string of a fixed length. Merkle network-based hash algorithms protect data integrity and authenticity by employing mathematical structures known as Merkle trees.

    \item [4. ] Elliptic isogenies-based digital signature schemes: Elliptic isogenies are mathematical transformations between elliptic curves. Elliptic isogeny-based digital signature techniques use the computational difficulties of generating inverse isogenies to guarantee message authenticity and non-repudiation.

    \item [5. ] Multivariate Cryptography: It is based on the use of multivariate equation systems to encrypt data and protect it from quantum attacks. Multivariate polynomials are utilized as encryption functions in a multivariate cryptography scheme. These polynomials have numerous variables and coefficients that are carefully chosen to protect the system's security. The computational complexity of solving systems of nonlinear multivariate equations is the strength of multivariate cryptography. Solving complex systems entails finding solutions that satisfy all the equations at the same time, which is a computationally challenging problem. Multivariate cryptography, on the other hand, poses difficulties. One of them is the size of the keys and signatures when compared to other cryptographic approaches. This can have an impact on system efficiency and performance. Furthermore, multivariate cryptography is susceptible to algebraic attacks and dimension-reduction techniques.

\end{itemize}

While post-quantum algorithms may require more computing resources and infrastructure compared to classical algorithms, which could lead to slower and more complex processes, they offer significant promise for enhancing information security in an increasingly interconnected world. The advent of quantum computers has exposed traditional cryptography to unprecedented vulnerabilities. However, post-quantum cryptography presents a viable solution to address these threats and ensure robust data protection.

\subsection{Related Works}

Stickel's key exchange protocol was motivated by the Diffie-Hellman protocol. In its original formulation, the protocol used the group of invertible matrices over a finite field~\cite{stickel2005new,myasnikov2008group}. Unfortunately, a linear algebra attack against this protocol has been published~\cite{shpilrain2008cryptanalysis,myasnikov2008group}. The attack is based on finding matrices \( x \) and \( y \) such that \( xa = ax \), \( yb = by \), and \( xu = y \), since \( x \) corresponds to \( a^{-n} \), while \( y \) equals \( b^m \)~\cite{grigoriev2014tropical}.

The Anshel-Anshel-Goldfeld algorithm defines a cryptographic primitive that utilizes non-commutative subgroups of a given platform group with efficiently computable normal forms. It was specifically implemented in the braid group. This scheme assumes that the Conjugacy Search Problem (CSP) is sufficiently difficult, which suggests it might also be implementable in other groups~\cite{myasnikov2008group}.

The Jintai Ding method leverages the Learning With Errors (LWE) and Ring-Learning With Errors (RLWE) problems to construct a post-quantum key exchange scheme. The fundamental idea behind this construction can be viewed as an extension of the Diffie-Hellman problem with errors~\cite{ding2012simple}, utilizing associativity and commutativity in a similar manner. The Kyber key establishment algorithm, selected by NIST, is a key encapsulation mechanism (KEM) based on public key encryption. In this scheme, Alice generates a ciphertext and a shared key using the input message, public key, and random vectors. Bob then uses the ciphertext and his private key to obtain the shared secret key. The security of this method relies on a variant of the NP-hard lattice problem known as learning with errors~\cite{nguyen2022analysis}.

Although quantum principles have posed threats to the security of major cryptographic systems~\cite{shor1994algorithms}, they have also led to the development of a new technology known as Quantum Key Distribution (QKD), which facilitates remote secret key establishment. QKD protocols exploit the principle that an eavesdropper cannot intercept or alter quantum communication without introducing detectable noise~\cite{bennett1984quantum}.

We will conclude this introduction by outlining the Diffie-Hellman algorithm, which played a crucial role in the development of public key cryptography. Additionally, we will discuss the Elgamal protocol, a public key cryptosystem that also provides a method for data encryption.

\subsection{Diffie-Hellman Key Exchange}

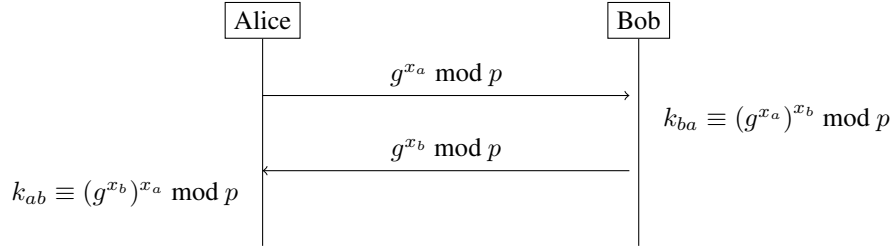
\begin{figure}
\centering
\begin{tikzpicture}
\def\ClientToServer{++(5,0)}
\def\ServerToClient{++(-5,0)}
\def\Lifeline{++(0,-3)}

\path (0,0) node[draw] (Alice) {Alice}
\ClientToServer node[draw] (Bob) {Bob};
\draw (Alice) -- \Lifeline (Bob) -- \Lifeline;

\path (Bob)
      ++(0,-1.0) node (BeginProcess) {} 
      node[below right] {${\hspace{2mm} k_{ba} \equiv (g^{x_a})}^{x_b} \ \text{mod} \ p$}
      ++(0,-1.0) node (EndProcess)   {};
(BeginProcess.west) rectangle (EndProcess.east);

\draw[->] (BeginProcess)\ServerToClient -- node[above] {$g^{x_a} \ \text{mod} \ p$} (BeginProcess);
\draw[->] (EndProcess) -- node[above] {$g^{x_b} \ \text{mod} \ p$} \ServerToClient;

\path (Alice)
      ++(0,-2.0) node (BeginProcess) {} 
      node[below left] {$k_{ab} \equiv (g^{x_b})^{x_a} \ \text{mod} \ p \hspace{2mm}$}
      ++(0,-2.0) node (EndProcess)   {};
(BeginProcess.west) rectangle (EndProcess.east);

\end{tikzpicture}
\vspace{-10mm}
\caption{Diffie-Hellman protocol: Both keys are identical ($k_{ab} \equiv k_{ba}$) because modular exponentiation follows the same rules as conventional exponentiation.}
\label{fig:Fig.1}
\end{figure}

The Diffie-Hellman (DH) key exchange~\cite{diffie1976new} operates in the field $\mathbb{Z}_p$, which is defined by two publicly shared parameters: the modulus $p$ and the generator $g$, which is a primitive root in $\mathbb{Z}_p$. Alice computes her public key $k_a \equiv g^{x_a} \mod p$ using a random integer $x_a$ as the exponent and sends it to Bob. Similarly, Bob computes his public key $k_b \equiv g^{x_b} \mod p$ and sends it to Alice. Each then performs exponentiation with the received value. Specifically, Alice computes ${(g^{x_b} \mod p)}^{x_a} \mod p \equiv g^{x_b x_a} \mod p$, and Bob computes ${(g^{x_a} \mod p)}^{x_b} \mod p \equiv g^{x_a x_b} \mod p$ (as illustrated in Fig.~\ref{fig:Fig.1}).

Because modular exponentiation adheres to the same laws as regular exponentiation, both computed values are identical. In the Diffie-Hellman algorithm, the prime number $p$ does not need to be kept secret. The security of the shared key relies on the difficulty of computing the value of $g^{x_a x_b} \mod p$ given $g$, $k_a$, and $k_b$.

\subsection{Elgamal Cryptosystem}

\begin{figure}
\centering
\begin{tikzpicture}
\def\ClientToServer{++(5,0)}
\def\ServerToClient{++(-5,0)}
\def\Lifeline{++(0,-3)}

\path (0,0) node[draw] (Alice) {Alice}
\ClientToServer node[draw] (Bob) {Bob};
\draw (Alice) -- \Lifeline (Bob) -- \Lifeline;

\path (Bob)
      ++(0,-1.0) node (BeginProcess) {} 
      node[below right] {$\hspace{2mm} k_{s} \equiv {(g^{y_a})}^{x_b} \ \text{mod} \ p$}
      ++(0,-1.0) node (EndProcess)   {};
(BeginProcess.west) rectangle (EndProcess.east);

\draw[->] (BeginProcess)\ServerToClient -- node[above] {$g^{y_a} \ \text{mod} \ p$} (BeginProcess);

\path (Alice)
      ++(0,-2) node (BeginProcess) {} 
      node[below left] {$k_{s} \equiv {(P_b)}^{y_a} \ \text{mod} \ p \hspace{2mm}$}
      ++(0,-2) node (EndProcess)   {};
(BeginProcess.west) rectangle (EndProcess.east);

\path (Bob)
      ++(0,-2.0) node (BeginProcess) {} 
      node[below right] {$\hspace{2mm} m \equiv m \cdot k_s \cdot {k_s}^{-1}  \ \text{mod} \ p$}
      ++(0,-2.0) node (EndProcess)   {};
(BeginProcess.west) rectangle (EndProcess.east);

\draw[->] (BeginProcess)\ServerToClient -- node[above] {$m \cdot k_s \ \text{mod} \ p$} (BeginProcess);

\end{tikzpicture}
\vspace{-10mm}
\caption{Elgamal cryptosystem: $P_b$ is Bob's public key, defined as $g^{x_b} \mod p$. Using \( y_a \) as a different random value for each message, Alice sends an encrypted message by multiplying \( m \) by \( k_s \). Bob retrieves $m$ by applying the inverse of $k_s$, denoted as ${k_s}^{-1}$. 
}
\label{fig:Fig.2}
\end{figure}
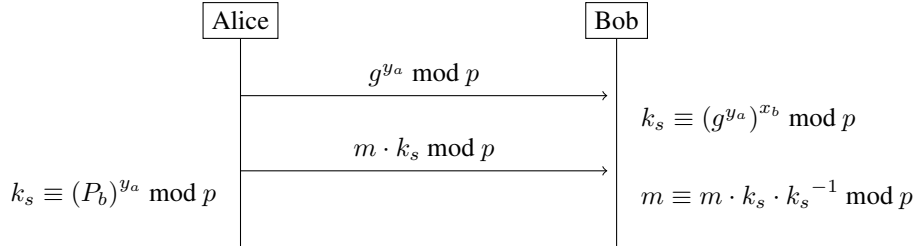

Since the Diffie-Hellman key exchange lacks user authentication, it is vulnerable to a Man In The Middle (MITM) attack. To address this issue, Elgamal proposed a variant of the protocol where users initially publish their public keys on a shared public platform~\cite{elgamal1985public}. In this variant, Bob's public key is computed as $P_b \equiv g^{x_b} \mod p$. When Alice wants to establish a secret key with Bob, she selects a random integer $y_a$ and sends $g^{y_a} \mod p$ to Bob, who then computes the shared key as $k_s \equiv (g^{y_a})^{x_b} \mod p$ (see Fig.~\ref{fig:Fig.2}). Alice computes $k_s \equiv (P_b)^{y_a} \equiv (g^{x_b})^{y_a} \mod p$ using Bob's public key $P_b$. Elgamal also proposes that a message $m$ can be encrypted or decrypted by multiplying or adding the multiplicative/additive inverse of $k_s$ in $\mathbb{Z}_p$.

The remainder of this article is organized as follows: Section 2 discusses the Key Establishment Method (KEM) and the associated security analysis. Section 3 examines the key sizes, while Section 4 explains Perfect Forward Secrecy (PFS). Finally, we provide our conclusions regarding the proposed cryptosystem.


\section{Key Establishment Algorithm}

In this novel key exchange algorithm, the public key of user \(i\) is the pair \((P_i, Q_i)\) defined within the finite field \(\mathbb{Z}_p\), where \(p\) is a publicly shared prime number. The public and private keys of Alice and Bob are presented in Table~\ref{tab:Table10}. The public keys \((P_i, Q_i)\) are computed according to Equation~\ref{eq:Eq.101}, where \(g\) is a primitive root modulo \(p\) publicly available.

\begin{equation}
\label{eq:Eq.101}
\begin{split}
P_i & \equiv {g}^{x_i + z_i} \ \text{mod} \ p\\
Q_i & \equiv {g}^{y_i + z_i} \ \text{mod} \ p
\end{split}
\end{equation}

\begin{table}[htbp]
\begin{center}
\caption{Key definitions for Alice and Bob in the proposed KEM (Key Establishment Method). The public key is the pair $(P_i, Q_i)$, while the private key consists of $(x_i, y_i, z_i)$. The generator $g$ is a publicly known primitive root of $p$.}
\begin{tabular}{|l|c|c|}
\hline
\textbf{User} & \textbf{Public Key} $(P_i, Q_i)$ & \textbf{Private Key} \\ \hline
\multirow{2}{*}{Alice} &
\multirow{2}{*}{$\left(g^{x_a + z_a}, \; g^{y_a + z_a}\right)$} &
\multirow{2}{*}{$(x_a, \; y_a, \; z_a)$} \\ 
& & \\ \hline
\multirow{2}{*}{Bob} &
\multirow{2}{*}{$\left(g^{x_b + z_b}, \; g^{y_b + z_b}\right)$} &
\multirow{2}{*}{$(x_b, \; y_b, \; z_b)$} \\ 
& & \\
\hline
\end{tabular}
\label{tab:Table10}
\end{center}
\end{table}

\begin{figure}
\centering
\begin{tikzpicture}
\def\ClientToServer{++(5,0)}
\def\ServerToClient{++(-5,0)}
\def\Lifeline{++(0,-3)}

\path (0,0) node[draw] (Alice) {Alice}
\ClientToServer node[draw] (Bob) {Bob};
\draw (Alice) -- \Lifeline (Bob) -- \Lifeline;

\path (Bob)
      ++(0,-0.5) node (BeginProcess) {} 
      node[below right] {${P_a}^{x_b}  {Q_a}^{y_b} \mod p$}
      ++(0,-0.5) node (EndProcess)   {};
(BeginProcess.west) rectangle (EndProcess.east);

\path (Bob)
      ++(0,-1.0) node (BeginProcess) {} 
      node[below right] {}
      ++(0,-1.0) node (EndProcess)   {};
(BeginProcess.west) rectangle (EndProcess.east);

\draw[->] (BeginProcess)\ServerToClient -- node[above] {$(P_a, Q_a)$} (BeginProcess);
\draw[->] (EndProcess) -- node[above] {$(P_b, Q_b)$} \ServerToClient;

\path (Alice)
      ++(0,-2) node (BeginProcess) {} 
      node[below left] {${P_b}^{x_a} {Q_b}^{y_a} \mod p \hspace{2mm}$}
      ++(0,-2) node (EndProcess)   {};
(BeginProcess.west) rectangle (EndProcess.east);

\path (Alice)
      ++(0,-2.5) node (BeginProcess) {} 
      node[below left] {}
      ++(0,-2.5) node (EndProcess)   {};
(BeginProcess.west) rectangle (EndProcess.east);

\end{tikzpicture}
\vspace{-20mm}
\caption{A public key exchange precedes key establishment.}
\label{fig:Fig.40}
\end{figure}
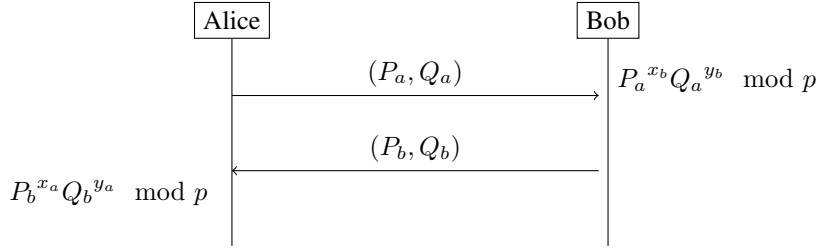

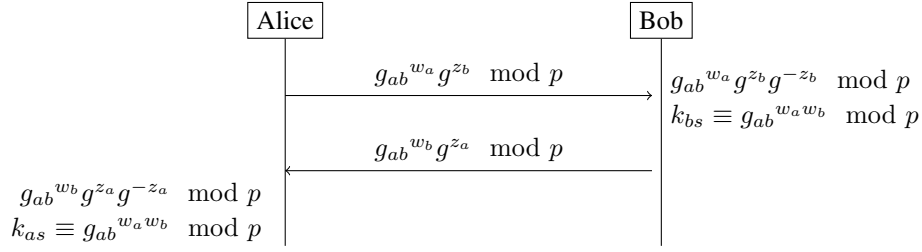
\begin{figure}
\centering
\begin{tikzpicture}
\def\ClientToServer{++(5,0)}
\def\ServerToClient{++(-5,0)}
\def\Lifeline{++(0,-3)}

\path (0,0) node[draw] (Alice) {Alice}
\ClientToServer node[draw] (Bob) {Bob};
\draw (Alice) -- \Lifeline (Bob) -- \Lifeline;

\path (Bob)
      ++(0,-0.5) node (BeginProcess) {} 
      node[below right] {${g_{ab}}^{{w_a}} g^{z_b} g^{-z_b} \mod p$}
      ++(0,-0.5) node (EndProcess)   {};
(BeginProcess.west) rectangle (EndProcess.east);

\path (Bob)
      ++(0,-1.0) node (BeginProcess) {} 
      node[below right] {$k_{bs} \equiv {g_{ab}}^{{w_a}{w_b}} \mod p$}
      ++(0,-1.0) node (EndProcess)   {};
(BeginProcess.west) rectangle (EndProcess.east);

\draw[->] (BeginProcess)\ServerToClient -- node[above] {${g_{ab}}^{{w_a}} g^{z_b} \mod p$} (BeginProcess);
\draw[->] (EndProcess) -- node[above] {${g_{ab}}^{{w_b}} g^{z_a} \mod p$} \ServerToClient;

\path (Alice)
      ++(0,-2) node (BeginProcess) {} 
      node[below left] {${g_{ab}}^{{w_b}} g^{z_a} g^{-z_a} \mod p\hspace{2mm}$}
      ++(0,-2) node (EndProcess)   {};
(BeginProcess.west) rectangle (EndProcess.east);

\path (Alice)
      ++(0,-2.5) node (BeginProcess) {} 
      node[below left] {$k_{as} \equiv {g_{ab}}^{{w_a}{w_b}} \mod p\hspace{2mm}$}
      ++(0,-2.5) node (EndProcess)   {};
(BeginProcess.west) rectangle (EndProcess.east);

\end{tikzpicture}
\vspace{-20mm}
\caption{Derivation of the secret key. Here, \( {g}_{ab}\) represents \( {g}^{x_a x_b + y_a y_b} \).}
\label{fig:Fig.401}
\end{figure}

\begin{itemize}
\item [1.] After Alice and Bob exchange their public keys (see Figure~\ref{fig:Fig.40}), each of them performs the computation specified by Equation~\ref{eq:Eq.20}.

\begin{equation}
\label{eq:Eq.20}
\begin{split}
{P_{i}}^{x_j}{Q_{i}}^{y_j}  & \equiv {\left( {g}^{x_i+z_i} \right) }^{x_j} { \left( {g}^{y_i+z_i} \right) }^{y_j} \ \text{mod} \ p \\
& \equiv {g}^{x_i x_j+y_i y_j}  {g}^{ z_i (x_j + y_j)}   \ \text{mod} \ p \\
& \equiv {g}^{x_i x_j+y_i y_j}  {g}^{ z_i (x_j + y_j)}   \ \text{mod} \ p \\
& \equiv {g}_{ij} {g}^{ z_i (x_j + y_j)}  \ \text{mod} \ p \\
\end{split}
\end{equation}

where ${g}_{ij}$ is used to represent ${g}^{x_i x_j+y_i y_j}$.

\item [2.] Each of them performs the computation specified by Equation~\ref{eq:Eq.2011}.

\begin{equation}
\label{eq:Eq.2011}
\begin{split}
\left( {g}_{ij} {g}^{z_i (x_j + y_j)} \right)^{{w_j}} & \equiv {{g}_{ij}}^{{w_j}} {g}^{z_i} \ \text{mod} \ p \\
\end{split}
\end{equation}

where \( w_j \) is the multiplicative inverse of \( x_j + y_j \) in \(\mathbb{Z}_{p-1}\), such that \( w_j (x_j + y_j) \equiv 1 \mod (p-1) \). Consequently, \( w_j \) is stated by Equation~\ref{eq:Eq.20202}.

\begin{equation}
\label{eq:Eq.20202}
w_j (x_j + y_j) = u_j (p-1) + 1 
\end{equation}

where \( u_j \) represents a variable within the system. However, we will treat \( w_j \) as a variable to avoid introducing this equation into the system. It is important to note that \( x_j + y_j \) must be invertible in \(\mathbb{Z}_{p-1}\). Then, Alice and Bob exchange their results (see Figure~\ref{fig:Fig.401}).

\item [3.] After reception, each participant multiplies by the inverse of \( g^{z_i} \) (or \( g^{z_j} \)), which results in \( g_{ij}^{w_j} g^{z_i} g^{-z_i} \) (or \( g_{ij}^{w_i} g^{z_j} g^{-z_j} \)). This simplifies to \( g_{ij}^{w_j} \) (or \( g_{ij}^{w_i} \)). Finally, to derive the secret key, each one performs exponentiation as specified by Equation~\ref{eq:Eq.2022}.

\begin{equation}
\label{eq:Eq.2022}
\begin{split}
k_{js} & \equiv \left( {{g}_{ij}} ^{{w_i}} \right) ^{{w_j}} \equiv  {g}^{ w_i w_j x_i x_j + w_i w_j y_i y_j} \ \text{mod} \ p \\
\end{split}
\end{equation}

Therefore, the keys are equal on both sides of the communication: \( k_{is} \equiv k_{js} \equiv k_{ij} \) (see also Table~\ref{tab:Table100}).

\end{itemize}

\subsection{Security Analysis}

\subsubsection{Public Key Analysis}

If Eve has access to a quantum computer, she can use the public keys of Alice, \(P_a \equiv {g}^{x_a + z_a} \mod p\) and \(Q_a \equiv {g}^{y_a + z_a} \mod p\) to determine \(x_a + z_a\) and \(y_a + z_a\). This results in a system of linear equations with two equations and three variables \((x_a, y_a, z_a)\), as shown in Equation~\ref{eq:Eq.35}, which is not solvable in general.

\begin{equation} \label{eq:Eq.35}
\begin{split}
x_a + z_a & = c_1 \\
y_a + z_a & = c_2 \\
\end{split}
\end{equation}

\subsubsection{Channel Analysis}

Now, let us assume that Eve is intercepting a message exchange between Alice and Bob, as illustrated in Figure~\ref{fig:Fig.401}. Eve's goal is to analyze the messages on the channel using her quantum computer. Consequently, she obtains \( w_a x_a x_b + w_a y_a y_b + z_b \) and \( w_b x_a x_b + w_b y_a y_b + z_a \). The system consists of 6 equations, as shown in Equation~\ref{eq:Eq.350}, involving 8 variables: \( x_a, y_a, z_a, w_a, x_b, y_b, z_b, w_b \).

\begin{equation} \label{eq:Eq.350}
\begin{split}
x_a + z_a & = c_1 \\
y_a + z_a & = c_2 \\
x_b + z_b & = c_3 \\
y_b + z_b & = c_4 \\
w_a x_a x_b + w_a y_a y_b + z_b & = c_5 \\
w_b x_a x_b + w_b y_a y_b + z_a & = c_6 \\
\end{split}
\end{equation}

where \( w_a \equiv (x_a + y_a)^{-1} \mod (p-1) \), and similarly for \( w_b \).

\subsubsection{Secret Key Analysis}

Suppose Bob, acting as a malicious Eve, first establishes a key with Alice, defined by Equation~\ref{eq:Eq.2022} as
\[
k_{ae} \equiv u^{({w_e}{w_a}) (x_a x_e + y_a y_e)} \mod p.
\] This scenario implies that Bob's variables are available to Eve, and the system of equations is represented in Equation~\ref{eq:Eq.351}.

\begin{equation} \label{eq:Eq.351}
\begin{split}
x_a + z_a & = c_1 \\
y_a + z_a & = c_2 \\
w_a x_a x_e + w_a y_a y_e & = c_3 \\
\end{split}
\end{equation}

where \( x_e, y_e \) are known by Eve, leaving a system of 3 equations with 4 variables: \( x_a, y_a, z_a, w_a \).

After conducting the previous analyses on Alice's public key, the exchange over the channel between Alice and Bob, and the scenario in which Bob acts maliciously, we can conclude that even with a quantum computer capable of computing discrete logarithms, Eve cannot derive Alice's private key.

Table~\ref{tab:Table100} provides a summary of the protocol that has been presented and discussed. It outlines each step involved in the process of securely deriving a shared secret key between Alice and Bob.

\begin{table}[h]
\caption{Steps of the protocol illustrating the sequence of operations (modulo $p$) leading to the derivation of the shared secret key by Alice.}
\centering
\begin{tabular}{|>{\centering\arraybackslash}m{6cm}|>{\setstretch{1.5}}m{8cm}|}
\hline
\textbf{Step} & \textbf{Description} \\ \hline
$g^{x_a} g^{z_a}, g^{y_a} g^{z_a}$ & Alice's public key components $(P_a, Q_a)$ sent to Bob \\ \hline

$\left( g^{x_a} g^{z_a} \right)^{x_b} \left( g^{y_a} g^{z_a} \right)^{y_b}$ & Bob computes exponentiation and multiplication of received components \\ \hline

$\left( g^{x_a x_b} g^{y_a y_b} g^{z_a (x_b + y_b)} \right)^{(x_b + y_b)^{-1}}$ & Bob raises the result to the inverse of $(x_b + y_b)$ \\ \hline

$g^{(x_b + y_b)^{-1} (x_a x_b + y_a y_b)} g^{z_a}$ & The intermediate result is sent from Bob to Alice \\ \hline

$\left( g^{(x_b + y_b)^{-1} (x_a x_b + y_a y_b)} g^{z_a} g^{-z_a} \right)^{(x_a + y_a)^{-1}}$ & Alice computes multiplication and exponentiation \\
$\equiv g^{(x_a + y_a)^{-1} (x_b + y_b)^{-1} (x_a x_b + y_a y_b)}$ & The final shared secret key \\ \hline

\end{tabular}
\label{tab:Table100}
\end{table}

\section{Key Sizes}

Let $|p|$ denote the length of the prime number $p$. Since each term of the public key $(P_i, Q_i)$ is computed in $\mathbb{Z}_p$, the size of the public key is $2 |p|$. 

The private key consists of $(x_i, y_i, z_i)$ each number chosen in the side of $|p|$. Consequently, the total size of the private key is $3|p|$. The sizes of the keys are summarized in Table~\ref{tab:Table200}.

\begin{table}[ht]
\centering
\caption{The size of the keys (represented in bits) as a function of $|p|$ for values of $|p|$ being 128, 256, and 512 bits. The table illustrates the key sizes for Private, Public, and Secret keys.}
\begin{tabular}{|l|c|c|c|}
\hline
\textbf{Key Type} & \textbf{128 bits} & \textbf{256 bits} & \textbf{512 bits} \\
\hline
Private: & 384 & 768 & 1536 \\
Public: $(P_i, Q_i)$ & 256 & 512 & 1024 \\
Secret: $(k_s)$ & 128 & 256 & 512 \\
\hline
\end{tabular}
\label{tab:Table200}
\end{table}

Table~\ref{tab:KeySizeComparison} presents a comparison of key sizes for three different cryptographic systems: Kyber512, ECDH (Elliptic-Curve Diffie-Hellman), and the proposed Key Establishment Method discussed in this work.

\begin{table}[ht]
\centering
\caption{Comparison of key sizes for Kyber512, ECDH, and the proposed KEM (Key Establishment Method).}
\begin{tabular}{|l|c|c|c|c|}
\hline
\textbf{KEM} & \textbf{Private Key} & \textbf{Public Key} & \textbf{Ciphertext} & \textbf{Secret Key} \\
\hline
Kyber512~\cite{avanzi2019crystals} & 1632 bytes & 800 bytes & 768 bytes & 256 bits \\
ECDH~\cite{ulla2023implementation} & 32 bytes & 64 bytes & -- & 256 bits \\
This Work & 96 bytes & 64 bytes & -- & 256 bits \\
\hline
\end{tabular}
\label{tab:KeySizeComparison}
\end{table}

\section{Perfect Forward Secrecy (PFS)}

Suppose that Alice and Bob renew their secret key periodically, and the current secret key \( k_{ab} \) has been compromised by an adversary. Perfect Forward Secrecy (PFS) is a security property of key agreement protocols that ensures the compromise of the current key does not undermine the security of previously established keys \( k_{0} \) and \( k_{1} \). According to Equation~\ref{eq:Eq.701}, Alice and Bob can generate a new secret key, \( k_{ab} \), by using the previously established keys \( k_{0} \) and \( k_{1} \).

\begin{equation}
\label{eq:Eq.701}
k_{ab} \equiv {g}^{k_{0}} \cdot {g}^{k_{1}} \ \text{mod} \ p
\end{equation}

For any given pair \(({g}^{k_{0}}, k_{ab}) \in \mathbb{Z}_p\), there always exists \({g}^{k_{1}}\) such that \(k_{ab} \equiv {g}^{k_{0}} \cdot {g}^{k_{1}} \mod p\). This is true because every integer in the ring \(\mathbb{Z}_p\) is invertible, which allows us to compute \({g}^{k_{1}}\) as \({g}^{k_{1}} \equiv k_{ab} \cdot {({g}^{-k_{0}})} \mod p\). Consequently, if only \(k_{ab}\) is known to the attacker, there are \(p-1\) possible pairs \(({g}^{k_{0}}_i, {g}^{k_{1}}_i)\) that satisfy the equation \(k_{ab} \equiv {g}^{k_{0}}_i \cdot {g}^{k_{1}}_i \mod p\), for \(1 \leq i < p-1\).

\section{Conclusions}
\label{Conclusions}

In this work, we have presented a novel algorithm for establishing a secret key between two remote entities using public keys. The security of our method is robust because it does not rely on the difficulty of integer factorization or the discrete logarithm problem.

After analyzing Alice's public key, the communication between Alice and Bob, and the scenario in which Bob acts like Eve, we can conclude that, even with a quantum computer capable of solving discrete logarithms, Eve is unable to obtain Alice's private key.

Moreover, our approach achieves Perfect Forward Secrecy (PFS), ensuring that the security of previously used keys remains intact even if a current key becomes compromised.

Additionally, the compact size of the public and private keys provided by our system makes it an attractive choice for practical implementations, particularly in scenarios where efficiency and minimal resource usage are critical.

\bibliographystyle{unsrt}

\end{document}